\documentclass[apjl]{emulateapj}
\usepackage{graphicx}
\usepackage{amssymb}
\usepackage{amsmath}
\usepackage{ulem}
\usepackage{soul}
\usepackage{threeparttable} 
\usepackage[colorlinks=true,linkcolor=blue,anchorcolor=blue,citecolor=blue,urlcolor=blue]{hyperref}  
\hypersetup{pdfauthor={Anna BARNACKA}, pdftitle={The size-duration relation for GRB progenitors}}

\def\aap{Astron.\ Astrophys.\ }
\def\mnras{Mon.\ Not.\ R.\ Astron.\ Soc.\ }
\def\apjl{Astrophys.\ J.\ Lett.\ }

\begin{document}

\title{A size-duration trend for gamma-ray burst progenitors}

\author{Anna~Barnacka}
\author{Abraham~Loeb}
\affiliation{Harvard-Smithsonian Center for Astrophysics, 60 Garden St, MS-20, Cambridge, MA 02138, USA}
\email[e-mails:~]{abarnacka@cfa.harvard.edu,aloeb@cfa.harvard.edu}

\begin{abstract}
Gamma-ray bursts (GRBs) show a bimodal distribution of  durations,
separated at a duration of $\sim$2~s.
Observations have confirmed the association of long GRBs 
with the collapse of massive stars. 
The origin of short GRBs is still being explored.
We examine  constraints on the size of emission region in short and long GRBs
detected by  {\it Fermi}/GBM. 
We find that the transverse extent of emission region during  the prompt phase, $R$, 
and the burst duration, $T_{90}$,
are consistent with the relation $R \sim c\times T_{90}$, for both long and short GRBs. 
We find the characteristic transverse extent for the prompt emission region to be 
$\sim 2 \times 10^{10}\,$cm, 
and $\sim  4 \times 10^{11}\,$cm  for short and long GRBs, respectively. 
\end{abstract}

\maketitle


\section{Introduction}
%

The first catalog of gamma-ray bursts (GRBs) detected by the BATSE satellite 
revealed bimodality  in their duration distribution \citep{1993ApJ...413L.101K}. 
Events  lasting longer than $2\,$s
were classified as  long GRBs,
while those shorter than $2\,$s were classified as short GRBs. 
This bimodal distribution suggested a difference 
in physical origin and  progenitor populations.

The discovery of optical afterglows following some of the long GRBs 
detected by {\it BeppoSAX} \citep{1997Natur.386..686V,1997Natur.387..878M}
and  environmental studies  indicated 
that long GRBs originate in star-forming galaxies \citep{1998ApJ...507L..25B,1998ApJ...508L..17D}
and their  location is spatially correlated with star-forming regions within their hosts \citep{2006Natur.441..463F}. 
In addition, some of the  long GRBs are associated with Type Ic supernovae \citep{2003Natur.423..847H,2003ApJ...591L..17S}.
These clues indicate that long GRBs are associated with the core collapse of massive stars (collapsars),
and not with the merger of  compact object binaries \citep{1998ApJ...494L..45P,1999ApJ...524..262M}.

The origin of short GRBs is still being explored. 
Short GRBs are often detected at lower redshift
 because these bursts are less energetic. 
They constitute less than 20\% of all detected GRBs. 
The lack of supernova associations shows that at least some 
short GRBs are not produced by massive star progenitors 
\citep{2005Natur.437..845F,2006ApJ...650..261S,2013ApJ...774L..23B}.

On theoretical grounds, the mergers of  compact binary objects 
have been investigated as a possible  origin of short bursts 
\citep{1998ApJ...494L..45P,1989Natur.340..126E,2006ApJ...648.1110B,2007ARep...51..308B}.
The compact object merger scenario is observationally supported by 
the location of short GRBs within their host galaxies 
\citep[for a recent review, see][]{2013arXiv1311.2603B}.

The transition between  short and long GRBs  at 2~s is 
somewhat arbitrary, and it is unclear how strongly it reflects differences in the progenitors of GRBs.
\citet{2013ApJ...764..179B} argued  that the transition between short and long GRBs is detector dependent,
and collapsars may be found among the short GRBs 
with some of the long GRBs produced by non-collapsars. 

The radius of the emission region and its transverse extent from the central engine 
is crucial in understanding the origin of gamma-ray emission,
and the nature of their progenitors 
\citep{2003ApJ...586..356Z,2008ApJ...675..519J,2009MNRAS.397..386Z,2010MNRAS.407.1033B,2010NewA...15..749T}.
Here we investigate the constraints on the size of the emission region and beaming factor 
of GRBs detected by {\it Fermi}/GBM \citep{2009ApJ...702..791M}. 
We examine the relation between  the transverse extent  of the prompt emission region and burst duration,
its implications for short and  long GRBs.

The paper is organized as follows.
In \S~\ref{sec:SourceSize} we introduce a formalism to constrain the size of the emission region.
Then, in \S~\ref{sec:DataAnalysis}, we describe the selection of GRBs detected by {\it Fermi}/GBM. 
The minimum variability time scales are estimated in \S~\ref{sec:minVar}.
The results are presented in \S~\ref{sec:results} and discussed in \S~\ref{sec:discussion}. 
Finally, we summarize our conclusions in \S~\ref{sec:summary}.


\section{Size of the emission region}
\label{sec:SourceSize}

The size of the prompt emission region is unresolved observationally. 
However, the light curve variability is attributed to the activity of the central engine \citep{1994ApJ...430L..93R,1997ApJ...485..270S},
and, as such, the minimum variability time scale provides a  constraint 
on the maximum transverse extent of the emission region (relative to the line-of-sight).
It is generally accepted that GRBs are powered by relativistic jets 
propagating with a speed $v=\beta c$ and Lorentz factor $\Gamma=(1-\beta^2)^{-1/2}$.
The maximum transverse extent  of the source is  given by the relation
\begin{eqnarray}
\label{eq:size}
R'_{max} &  \simeq & c\,t'_{var} \simeq\frac{\mathcal{D}\,c\,t_{var}\,}{(1+z)} \\ 
\nonumber                 &  \simeq & \frac{3\times10^{11}\,\mbox{cm}}{(1+z)} \left( \frac{\mathcal{D}}{100}\right)  \left( \frac{t_{var}}{0.1\,\mbox{s}}\right)  \,, 
\end{eqnarray}
where prime denotes quantities in the comoving  frame of the emitting plasma, 
$t_{var}$ is the observed minimum variability time scale,  $z$ is the cosmological redshift,
and $\mathcal{D}=[\Gamma(1-\beta\cos\theta_{\rm obs})]^{-1}$ is the Doppler factor of the observed radiation,
with $\theta_{\rm obs}$ being the observer viewing angle relative to the velocity of the emitting material. 
For typical GRBs, 
$\theta_{\rm obs}\lesssim1/\Gamma$ and $\mathcal{D} \sim \Gamma$.

Setting constraints on the maximum radius of the emission region requires knowledge of the Doppler factor $\mathcal{D}$. 
There  are several methods for placing lower and upper limits on $\mathcal{D}$ 
\citep{1993ApJ...409..327B,1997ApJ...491..663B,2001ApJ...555..540L,2004ApJ...613.1072R,2011ApJ...738..138R},
which provide values in the range  20 - 3000.

The emission region has to be transparent to gamma rays.
A small radius for the emission region  and high photon densities 
imply a very large optical depth to $\gamma\gamma$ absorption and pair creation \citep{1991ApJ...373..277K,1997ApJ...491..663B}.
In order to allow gamma-ray photons to escape from the emission region,
the optical depth for $\gamma\gamma$ absorption, $\tau_{\gamma\gamma} $ \citep{1967PhRv..155.1408G}, 
must satisfy 
\begin{equation}
\label{eq:taugg}
\tau_{\gamma\gamma}(E) \simeq R' \int_{-1}^{1} d\mu \frac{1-\mu}{2} \int_{E_{th}}^{\infty} d\epsilon\, n(\epsilon) \sigma_{\gamma\gamma}  < 1 \,,
\end{equation}
assuming isotropic emission in the comoving frame,
where $R'$ is a transverse extent  of the emission region, $\mu=\cos\theta$, 
where $\theta$ is an angle between the momenta of the emitted photon and the ambient photon, 
and $\sigma_{\gamma\gamma} (E,\epsilon,\mu)$  is the polarization averaged 
cross-section for pair production \citep{1976tper.book.....J,2014arXiv1409.1674B}. 
Here, $E$ is the energy of the emitted photon,  $\epsilon$ is the energy of the ambient photon,
and $E_{th}=2/(\epsilon(1-\mu))$ is the threshold energy for  pair production.
All photon energies are defined in the jet comoving frame and  in dimensionless  units, normalized by  $m_ec^2$.

The spectral photon number density is given by
\begin{equation}
\label{eq:ne}
n(\epsilon) = \frac{d_L^2\,\Phi(\epsilon)}{\mathcal{D}^4\,R'^2\,m_ec^3\epsilon}  \,,
\end{equation}
where $d_L$ is the luminosity distance to the source, $\Phi$ is the observed  energy flux (also known as $\nu F_{\nu}$).
Combining equations (\ref{eq:taugg}) and (\ref{eq:ne}), 
 assuming the transverse extent of the emission region to be $R'_{max}$, gives the minimum Doppler factor
\begin{equation}
\label{eq:Doppler}
\mathcal{D}^5  > \frac{(1+z)d_L^2}{c\, t_{var}}\, \Upsilon \,,
\end{equation} 
where
\begin{equation}
\Upsilon = \int_{-1}^{1} d\mu \frac{1-\mu}{2} \int_{E_{th}}^{\infty} d\epsilon \, \frac{\Phi(\epsilon)}{m_ec^3\epsilon} \sigma_{\gamma\gamma} \,.
\end{equation}

The minimum Lorentz  factor, $\Gamma_{min}$, required for gamma rays to escape 
from the emission region depends on the combination of the emitted and  ambient photon energies. 
Conservatively we evaluate $\Gamma_{min}$ at  an energy  $E=2\,E_{th}$,
where the  $\gamma\gamma$ absorption cross-section has a distinct maximum.  
The threshold energy is calculated using the ambient photon energy, $\epsilon$, 
equal to the peak energy of the spectral energy distribution of the prompt emission, $E_{p}$. 

Equations (\ref{eq:taugg}) and (\ref{eq:ne}) can also be used to evaluate the minimum transverse extent of the emission region
\begin{equation}
R_{min}'  \simeq \frac{d_{L}^2}{\mathcal{D}^4}  \, \Upsilon  \,,
\label{eq:Rmin}
\end{equation}
assuming conservatively a maximum  value of $\mathcal{D}\sim$1200 \citep{2011ApJ...738..138R}.

\section{Data Selection}
\label{sec:DataAnalysis}

The Gamma-Ray Burst detector (GBM)\citep{2009ApJ...702..791M} on board the Fermi satellite 
consists of 12 NaI and 2 BGO scintillators 
which cover the energy range from 8 keV - 40 MeV, in 128 energy bins,
and monitor the entire sky. 
In the five years of its operation, the GBM triggered on more than 1300 GRBs.

Information on the GRB distances is required for obtaining  robust limits on the emission region sizes.
We have therefore selected a limited sample of $\sim50$ GRBs with measured redshifts.
Of those, only bursts with good spectral quality and light curves with sufficient signal-to-noise ratio were kept. 
The primary condition for the data selection
was at least one light curve with signal-to-noise ratios greater than~2 at a sampling of 0.128$\,$s.
The selection procedure reduced the initial sample to 24~long GRBs with measured redshifts. 

To add short GRBs,
we have selected GRBs detected by the GBM 
with duration $T_{90}$\footnote{$T_{90}$ is defined as the time interval 
over which 90\% of the flux, integrated over the burst duration, was detected. 
(http://heasarc.gsfc.nasa.gov/W3Browse/fermi/fermigbrst.html\#t90)  } 
shorter than 2~s, 
and a flux greater than 15~photons~cm$^{-2}$s$^{-1}$.  
The flux was measured in the energy range 10-1000 keV with 64~ms sampling time.  
Altogether, we have selected 43~short GRBs.
The quality selection,  the same as for long GRBs, limited the sample to 19~short GRBs.
The entire sample of short GRBs has unknown redshifts,
and so we have assumed an average redshift of 0.85 \citep{2014arXiv1405.5131D}.

The light curve analysis and a spectra fitting in our analysis have been performed following the procedures described by Holland et al. (2012)
\footnote{http://fermi.gsfc.nasa.gov/ssc/data/analysis/user/do\_gbm.pdf}.


\begin{figure*}
\label{fig:lc} 
\includegraphics[width=19cm]{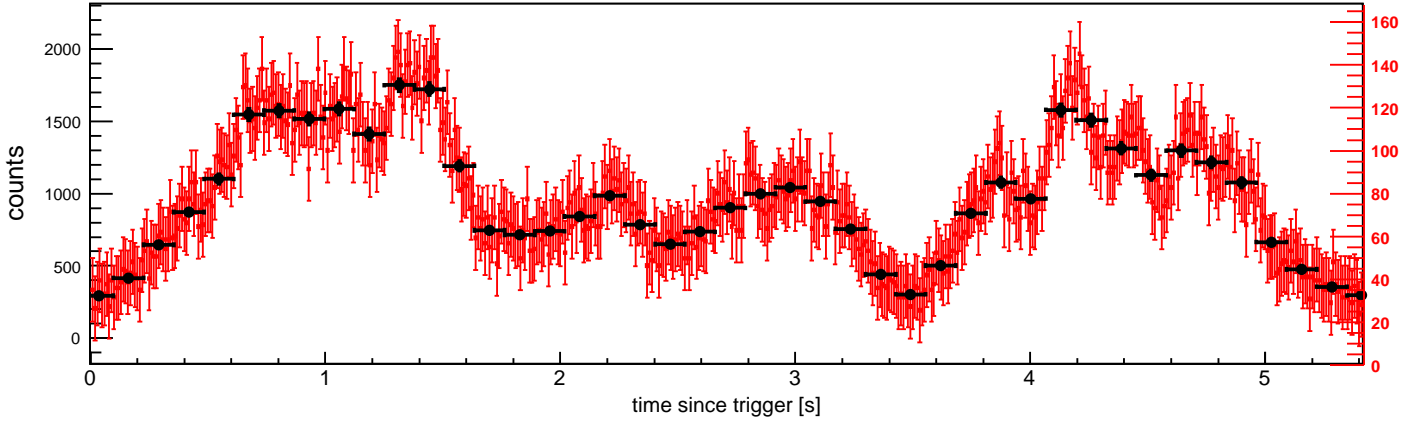}
\caption{Sample light curve of GRB090424  using two binnings;  0.128~s represented by black points, and 0.01 by red points.}
\end{figure*}

\section{Minimum Variability Timescale}
\label{sec:minVar}

The Fermi/GBM triggers on bursts within 16 ms. 
All triggers generate time-tagged event data (TTE) consisting of the photon arrival
time and energy as deposited in each of the 14 detectors with a temporal resolution of
2 $\mu$s \citep{2009ApJ...702..791M}. 

The intrinsic minimum variability time scale of GRBs is determined  by 
the size of  the emission region and the emission mechanism. 
The observed minimum variability time scales are, in addition, limited by 
the sensitivity of the detectors, the GRBs flux and the sampling of light curves. 
Figure~\ref{fig:lc} shows the light curve of  one of the brightest gamma-ray bursts detected by GBM: GRB090424. 
The light curve is displayed with two different binnings, $0.128\,$s and $0.01\,$s,
demonstrating how  binning can smooth variability, 
or   decrease signal-to-noise ratio. 

To find the observed minimum variability time scale for each GRB, we have used the method 
utilized by \citet{2012ApJ...744..141B,2013EAS....61...45B,2013arXiv1307.7618B},
which searches for a characteristic time scale at which the variance ratios per bin width is minimum.
The characteristic time scale is interpreted as an upper limit on the minimum variability time scale.
This method incorporates  the following steps:
first, the time interval of the prompt emission is selected based on  $T_{90}$; 
then, a background time interval of an equal  duration is selected. 
Both the signal and the background intervals are used to derive differentials,
which, in the next step, are used to calculate variances of  the signal and the background. 
The ratios of the variances are calculated for different binnings
in the range from $10^{-3}\,$s up to 0.1$\times T_{90}$ using ten logarithmic bins per decade.
The bin width at which the variance ratio divided by the bin width obtains its minimum value is interpreted 
as a minimum observed variability time scale, $t_{var}$, \citep[e.g., see Figure~1 in][]{2013EAS....61...45B}.
The resulting minimum variability time scales for the entire sample are listed in Table~\ref{tab:short}.

The variance ratio contains information on the rates of change in the light curves.
These rates of change can decrease either because of the lack of  further variability of the sources, or due to limited photon statistics. 

To assess whether the observed minimum variability time scale can be intrinsic to the source, 
we have calculated an average significance of the signal in the light curve for each binning, $\bar{N_\sigma}$.
If the average significance at the observed minimum variability time scale is $>4\, \sigma$,  
the observed variability time scale is interpreted as the intrinsic property of the emission region.
 
The average significance, $\bar{N_\sigma}$, 
of the light curve with a binning corresponding to $t_{var}$  
is listed in Table~\ref{tab:short} for samples of long and short GRBs.   

We have found that the average minimum variability time scale obtained for short GRBs is $0.036\,$s,
and $1.2\,$~s for long GRBs. 
In our samples, the observed variability time scales for long GRBs are significantly longer than those of short GRBs. 
This is in agreement with the minimum variability 
timescales of long and short GRBs obtained by \citet{2013MNRAS.432..857M}.

The high significance  obtained at $t_{var}$ suggests that 
for the majority of the long GRBs in our sample, the observed variability time scale 
is not limited by the statistics. 
In the case of short GRBs, the average significance at the observed minimum variability time scale is small,
suggesting that the intrinsic variability time scales for these GRBs can be shorter  than  observed.

\citet{2013MNRAS.432..857M}  noticed a correlation between 
the minimum variability timescale and the burst duration for short GRBs.
A hint of such a correlation is also present in our sample. 
However, as we have shown, the average significance at
 the characteristic variability timescale is small
for short GRBs.
Thus, the correlation can be the result  of limited photon number statistics. 

In our sample, the observed minimum variability time scales for long gamma-ray bursts 
does not show a correlation with the duration of  the burst, $T_{90}$.
This is in agreement with \citet{2014ApJ...787...90G}, 
who investigated the minimum variability timescales for a large sample of {\it Swift} GRBs.

\begin{figure}[h!]
\begin{center}
\includegraphics[width=9.5cm,angle=0]{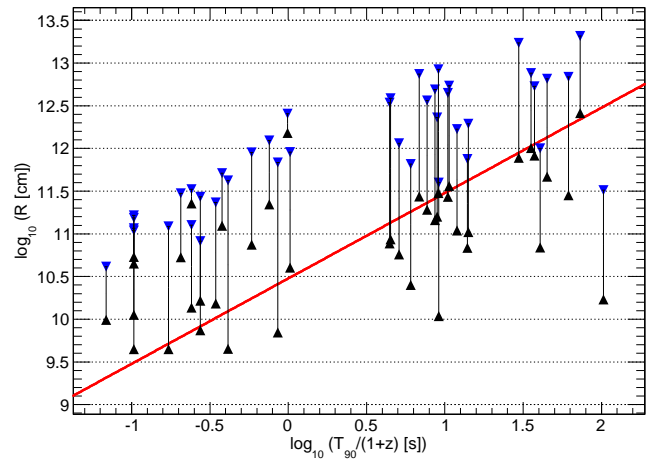}
\end{center}
\caption{\label{fig:grbsize} Upper and lower limits on the transverse extent of the emission region as a function of $T_{90}/$(1+z).
						       Blue down-pointing triangles indicates upper limits,
						      and black up-pointing triangles corresponds to lower limits. 
						     The red line equates the transverse extent  of the emission region to $c\times T_{90}/$(1+z),
						      where $T_{90}/$(1+z) is an engine lifetime.
						      Note that this line is not attempting to fit the constraints from the data}. 
\end{figure}

\begin{figure}
\begin{center}
\includegraphics[width=9.5cm,angle=0]{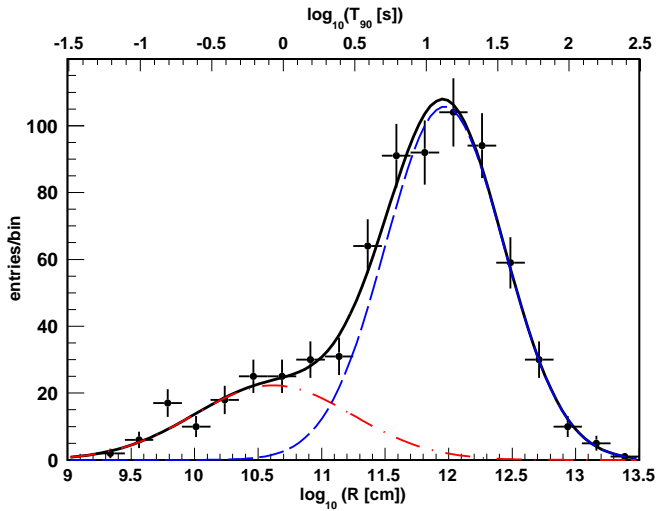}
\end{center}
\caption{\label{fig:size_distribution} Size distribution of emission regions  during the prompt emission of GRBs.
						      The distribution is obtained for a total of 1349 GRBs detected by {\it Fermi}/GBM.
						      The transverse extents  are estimated using the duration $T_{90}$ multiplied by the speed of light,
						       without correcting for the (1+ z) redshift factor. 
						      The solid line delineates a  fit to the distribution involving two Gaussian functions (in blue and red).}
\end{figure}


\section{Results}
\label{sec:results}

Following the formalism described in  \S~\ref{sec:SourceSize},
we have obtained the maximum and minimum transverse extents  of the prompt emission regions 
for a sample of 19 short and 24 long GRBs. 
Table~\ref{tab:short}  lists the properties for each burst.

Figure~\ref{fig:grbsize} shows  upper and lower limits on the transverse extent  of the emission region
 as a function  of the duration of the engine activity $T_{90}/(1+z)$. 
The upper limits were obtained using equation~(\ref{eq:size}), 
where the minimum variability time scale was derived from 
the light curve analysis described in \S~\ref{sec:minVar},
and the Doppler factor was constrained using equation~(\ref{eq:Doppler}).
The energy flux, $\Phi(\epsilon)$, 
was obtained by fitting the Band function~\citep{1993ApJ...413..281B} to the spectrum. 

\section{Discussion}
\label{sec:discussion}

Our  constraints on the transverse extent  of prompt emission region is based on the analysis of the minimum variability time scales 
and the Doppler factors, which are evaluated using $T_{90}$ and the same energy range for all GRBs.

The luminosity distances in the sample of long GRBs have been calculated using measured redshifts,
whereas  for the sample of short GRBs we have assumed an average redshift of 0.85 \citep{2014arXiv1405.5131D}. 
The constraints on the Doppler factor scale with the luminosity distance as $\mathcal{D}\,\alpha \, d_{L}^{2/5}$.
Even if short GRBs are located  5 times  farther away, 
the observed size would be underestimated by merely a factor of 2. 
The maximum transverse extent  of the emission regions 
of long GRBs is larger than that of short GRBs by over an order of magnitude. 
Thus, the assumed redshift for short GRBs has little impact on our results. 

The  minimum transverse extent  of the emission region has been estimated differently for short and long GRBs.
For the majority of the  long GRBs, the observed minimum variability time scale 
is not limited by the photon counting statistics, and thus may 
refer to the intrinsic variability of the source. 
Therefore, the minimum transverse extent  of the emission regions has been calculated as $\Gamma_{min}\,c\, t_{var}/(1+z)$,
assuming conservatively a minimum value of Lorenz factor, $\Gamma_{min}$, 
$\sim$20 \citep{2012MNRAS.420..483G,2014arXiv1408.3042S}.

The observed variability time scale for short GRBs is limited by  photon statistics; 
thus the emission region can be  smaller than estimated. 
To evaluate the minimum transverse extent  of the emission region in the sample of short GRBs, 
we have used equation~(\ref{eq:Rmin}). 

The emission region of GRBs is expected to expand close to the speed of light.
We have therefore compared our results to $c\times T'_{90}$.
Remarkably, this relation lies  within the constrained transverse extent  of the emission region of the sample of long and short GRBs. 

Figure~\ref{fig:size_distribution} shows the transverse extents  distribution of the emission regions of all GRBs detected by the GBM detector until June 2014.
The transverse extents  have been estimated using the relation $c\times T_{90}$.  
The distribution of transverse extents  is bimodal, 
with an average transverse extent  of the emission regions of $\sim 2 \times 10^{10}\,$cm, 
and $\sim 4 \times 10^{11}\,$ cm, for short and  long GRBs, respectively.
These transverse extents are corrected for the redshift factor assuming the average redshift of z=0.85 \citep{2014arXiv1405.5131D} for short GRBs,
and  the average redshift of z=1.77 for long GRBs listed in Table~\ref{tab:short}.


\section{Summary}
\label{sec:summary}

The bimodality in the distribution of GRB durations has been interpreted as 
evidence for  two  progenitor populations. 
In the collapsar scenario, the central engine has to be active long enough for the emission region to exit  
the stellar envelope and produce the observed $\gamma-$rays.
In this scenario, the radius of the stellar envelope is $\lesssim10^{12}\,$cm.

Our constraints imply that the emission region size during  the prompt GRB emission, $R$, 
and the central engine duration, $T_{90}$,
 are consistent with the relation $R \sim c\times T_{90}$. 
We have obtained the characteristic transverse extent of the prompt emission region to be $\sim 2 \times 10^{10}\,$cm, 
and $\sim 4 \times 10^{11}\,$~cm for short and long GRBs, respectively. 

The average beaming factor for short GRBs in our sample is 670,
yielding a radius of emission region, $r\sim\Gamma R$, of $\sim10^{13}\,$cm. 
The average beaming factor for long GRBs is 450,
implying  the radius of emission region of $\sim 2\times 10^{14}\,$cm. 

\begin{table*}[H] 
 \centering  
 \begin{threeparttable} 
 \caption{\label{tab:short}  Sample of short and long GRBs.   } 
\begin{ruledtabular}
 
 \begin{tabular}{cccccccccc}
Name  &  Redshift & $T_{90}$  & $T_{50}$ & $t_{var}$ & $\bar{N_\sigma}$ &  $\mathcal{D}_{min}$ & $R_{min}$ & $R_{max}$   \\
             & (z)    &    [s]         & [s]     & [s]            &    &                                   &        [cm]           &   [cm]   \\
\hline
GRB081209981 & -- & 0.192  & 0.128  & 0.011  & 0.3 &  230  & 1.3e+09 & 3.9e+10 \\                                                                                                 
GRB081216531 & -- & 0.768  & 0.128  & 0.067  & 0.4 &  390  & 4.5e+09 & 4.2e+11 \\ 
GRB090108020 & -- &0.704   & 0.256  & 0.038 & 0.2 &  850  & 1.2e+11 & 5.1e+11 \\ 
GRB090228204  & -- & 0.448 & 0.128  & 0.011 & 0.4 &  700  & 1.4e+10 & 1.2e+11 \\ 
GRB090328713 & -- &0.18     & 0.128  & 0.011 & 0.2 &  680     & 1.1e+10 & 5.1e+11 \\ 
GRB090802235 & -- &0.128   & 0.064  & 0.003 & 0.4 &  850  & 9.8e+09 & 4.1e+10 \\ 
GRB100929916  & -- &0.320  & 0.256  & 0.014 & 0.1 &  530  & 4.4e+09 & 1.2e+11 \\ 
GRB101216721  & -- & 1.917 & 0.512  & 0.100 & 0.4 &  560  &  4.0e+10 & 9.0e+11 \\ 
GRB110705151 & -- & 0.17    & 0.128  & 0.011 & 0.6 &  910    & 5.4e+10 & 1.6e+11 \\ 
GRB111112908 & -- & 0.192  & 0.128  & 0.011 & 0.2 &  890  &  4.5e+10 & 1.5e+11 \\ 
GRB120222021 & -- & 1.088  & 0.512  & 0.086 &  1.1 &  650  &  7.4e+10 & 8.9e+11 \\ 
GRB120323507 & -- & 0.448  & 0.192  & 0.019 & 3.3 &  1100  &  2.3e+11 & 3.3e+11 \\ 
GRB120624309 & -- & 0.640  & 0.160  & 0.024 & 0.8 &  610  &  1.5e+10 & 1.1e+11 \\ 
GRB130416770 & -- & 0.19    & 0.048  & 0.012 & 0.2 &  550  & 4.4e+09 & 1.1e+11 \\ 
GRB130504314 & -- & 0.384  & 0.192  & 0.024 & 0.6 &  780  &  5.3e+10 & 2.9e+11 \\ 
GRB130628860 & -- & 0.512  & 0.384  & 0.008 & 0.2 &  670  & 7.4e+09 & 8.2e+10 \\ 
GRB130701761 & -- & 1.600  & 0.704  & 0.109 & 3.5 &  390  & 7.0e+09 & 6.8e+11 \\ 
GRB130912358 & -- & 0.512  & 0.192  & 0.028 & 2.0 &  600  & 1.6e+10 & 2.7e+11 \\ 
GRB140209313 & -- & 1.408  & 0.320  & 0.099 & 1.3 &  780  & 2.2e+11 & 1.2e+12 \\ 
\hline
GRB080804972 & 2.205 & 24.704 & 10.432 &1.021 & 6.3   &  380   & 3.1e+10 & 3.6e+12 \\ 
GRB080916009 & 0.689 &  62.97  & 32.000 & 2.323 & 10.2 &  130   & 7.0e+10 & 5.4e+12 \\ 
GRB080916406 & 4.350 & 46.337 & 18.432 & 1.291 & 2.1   &  680   & 3.9e+10 & 4.9e+12 \\ 
GRB081121858 & 2.512 & 41.985 & 9.472 & 0.638 & 0.5   &  310   & 1.9e+10 & 1.7e+12 \\ 
GRB081221681 & 2.260 & 29.697 & 7.488 & 1.633 &  8.6  &  560   & 4.9e+10 & 8.4e+12 \\ 
GRB081222204 & 2.700 & 18.880 & 4.672 & 0.355 & 11.3 & 400    & 1.1e+10 & 1.1e+12 \\ 
GRB090102122 & 1.547 & 26.624 & 9.728 & 1.148 & 9.3   &  330   & 3.4e+10 & 4.5e+12 \\ 
GRB090323002 & 3.570 & 135.17 & 53.249 & 5.943 & 11.0 &  440   & 1.7e+11 & 1.7e+13 \\ 
GRB090328401 & 0.736 & 61.697 & 14.592 & 0.3      & 4.5   &  150   & 8.8e+10 & 7.6e+12 \\ 
GRB090424592 & 0.544 & 14.144 & 3.072 & 0.028 & 7.0   &  730   & 8.4e+08 & 3.9e+11 \\ 
GRB090618353 & 0.540 & 112.38 & 23.808 & 6.683 & 3.0   &  160   & 2.0e+11 & 2.1e+13 \\ 
GRB090902462 & 1.822 & 19.328 & 9.024 & 1.291 & 7.8   &  540   & 3.9e+10 & 7.4e+12 \\ 
GRB090926181 & 2.106 & 13.760 & 6.528 & 0.399 & 8.8   &  890   & 1.2e+10 & 3.4e+12 \\ 
GRB091003191 & 0.897 & 20.224 & 13.312 & 1.148 &  3.1  &  300   & 3.4e+10 & 5.4e+12 \\ 
GRB091020900 & 1.710 & 24.256 & 6.912 & 0.718 & 0.5   &  290   & 2.1e+10 & 2.3e+12 \\ 
GRB091024372 & 1.092 & 93.954 & 39.937 & 1.633 &  1.5  &  280   & 4.9e+10 & 6.5e+12 \\ 
GRB091208410 & 1.063 & 12.480 & 7.168 & 0.087 &  0.5  &  520   & 2.6e+09 & 6.5e+11 \\ 
GRB100814160 & 1.440 & 150.53 & 72.193 & 1.148 &  12.0&  490   & 3.4e+10 & 6.9e+12 \\ 
GRB100906576 & 1.727 & 110.59 & 18.944 & 0.316 & 7.6   &  290   & 9.5e+09 & 1.0e+12 \\ 
GRB110213220 & 1.460 & 34.305 & 6.400 & 0.281 & 3.5   &  220   & 8.4e+09 & 7.5e+11 \\ 
GRB120712571 & 4.000 & 22.528 & 7.424 & 0.718 &  0.5  &  900   & 2.1e+10 & 3.9e+12 \\ 
GRB120729456 & 0.800 & 25.472 & 8.320 & 0.316 &  3.0  &  370   & 9.5e+09 & 1.9e+12 \\ 
GRB130427324 & 0.340 &138.24  & 4.096 & 0.038 &  7.5  &  380   & 1.1e+09  & 3.2e+11 \\                                                                                          
GRB130518551 & 2.490 & 3.456   & 1.280 & 0.280  &  3.1  &  1060 & 1.5e+12 & 2.5e+12 \\      
 \end{tabular}
 
 \begin{tablenotes}
      \small
      \item Note: column labels denote the
					names of the GRBs,  GRB duration ($T_{90}$),
					minimum variability time scale ($t_{var}$), 
					the average significance of the signal in the light curve with binning corresponding to $t_{var}$,
					the minimum Doppler factor ($\mathcal{D}_{min}$),
					the minimum size of the emission region ($R_{min}$),
					and the maximum size of the emission region  ($R_{max}$).
    \end{tablenotes}
\end{ruledtabular}
\end{threeparttable} 
\end{table*}

\acknowledgments

We thank the referee for valuable comments on the manuscript.
We thank Markus B\"ottcher,  Edo Berger, Josh Grindlay, and Raffaella Margutti
for comments on the manuscript and for useful discussions.
The work of A.B. is supported by the Department of Energy Office of Science, NASA \& the
Smithsonian Astrophysical Observatory and financial support by the NCN grant DEC-2011/01/M/ST9/01891
is acknowledged.
This work was also supported in part by NSF grant AST-1312034.


\end{document}